\documentclass[aps,prb,twocolumn,superscriptaddress,showpacs,superbib,floats]{revtex4}
\usepackage{graphicx}
\usepackage{amssymb}
\usepackage{amsmath}
\usepackage{dcolumn}
\usepackage{color}

\begin{document}

\title{Electronic and phononic properties of the chalcopyrite CuGaS$_2$}

\author{A.H. Romero}
\affiliation{CINVESTAV, Departamento de Materiales, Unidad
Quer$\acute{e}$taro, Quer$\acute{e}$taro, 76230, Mexico}

\author{M. Cardona}
\author{R. K. Kremer}
\email[Corresponding author:~E-mail~]{R.Kremer@fkf.mpg.de}
\author{R. Lauck}
\author{G. Siegle}
\author{C. Hoch}
\affiliation{Max-Planck-Institut f{\"u}r Festk{\"o}rperforschung,
Heisenbergstrasse 1, D-70569 Stuttgart, Germany}

\author{A. Mu\~{n}oz}
\affiliation{MALTA Consolider Team, Departamento de F\'{\i}sica Fundamental II, and Instituto de Materiales y Nanotecnolog\'{\i}a, Universidad de La Laguna, La Laguna 38205, Tenerife, Spain}

\author{A. Schindler}
\affiliation{NETZSCH-Ger\"atebau GmbH, Wittelsbacherstr. 42, D-95100 Selb, Germany}

\date{\today}

\begin{abstract}
The availability of \textit{ab initio} electronic calculations and the
concomitant techniques for deriving the corresponding lattice dynamics
have been profusely used  for calculating thermodynamic
and vibrational properties of semiconductors, as well as their
dependence on isotopic masses. The latter have been compared with experimental data for elemental and binary semiconductors with different
isotopic compositions. Here we present theoretical and experimental
data for several vibronic and thermodynamic properties of CuGaS$_2$, a canonical
ternary semiconductor of the chalcopyrite family. Among
these properties are the lattice parameters, the phonon dispersion relations
and densities of states (projected on the Cu, Ga, and S constituents),
the specific heat and the volume thermal expansion coefficient. The
calculations were performed with the ABINIT and VASP codes within
the LDA approximation for exchange and correlation and the results are compared with data obtained on samples with the natural isotope composition for Cu, Ga and S, as well as for isotope enriched samples.

\end{abstract}

 \maketitle


\email{R.Kremer@fkf.mpg.de}

\section{Introduction}\label{SecIntro}

The availability of \textit{ab initio} electronic calculations and the
concomitant techniques for deriving the corresponding lattice dynamics
have been profusely used in the past decade for calculating thermodynamic
and vibrational properties of semiconductors, as well as their
dependence on isotopic masses. The latter have been compared with experimental data for elemental and binary semiconductors with different
isotopic compositions.\cite{Sanati2004,Gibin2005,Cardona2010a} Here we present theoretical and experimental
data for several vibronic and thermodynamic properties of a canonical
ternary semiconductor of the chalcopyrite family: CuGaS$_2$.
Two main groups of chalcopyrites are usually considered, one, denoted as I-III-VI$_2$, derived from the II-VI  compounds with zincblende structure, the other, II-IV-V$_2$ , derived from the III-V zincblende compounds. Examples of the first group are CuGaS$_2$ and AgGaS$_2$, whereas the second type is represented, for example, by ZnGeAs$_2$. The chalcopyrite structure has the space group $I$$\bar{4}$2$d$ and  the class $\bar{4}$2$m$ with two formula units per primitive cell and a longitudinal distortion along the $c$-axis which converts the tetrahedral primitive cell (PC) into a tetragonal one.  The lattice constants of the tetragonal PC are $a$ = $b$  (along $x$ and $y$) and $c$ (along $z$).\cite{Jaffe1984}   The regular tetrahedra with the anion at the center and the cations at the vertices are distorted because,
e.g., of the different lengths of the I-II  and the III-II bonds.  The anion distortion is usually chosen to be along the $x$ direction and equal to $u$ - 1/4,  $u$ = 1/4  corresponds to no tetrahedral distortion.
In this article we discuss lattice parameters and vibrational properties of some I-III-VI$_2$  chalcopyrites (I = Cu, Ag). The motivation for this choice (as opposed to the II-VI-V$_2$ materials) is that the copper and silver chalcopyrites have received considerable attention for the production of photovoltaic cells. Their energy gaps cover the range 1 - 3.5 eV, i.e., most of the frequency of the solar spectrum. From the fundamental point of view, these materials have the property that the 3$d$ core electrons of Cu and the 4$d$ of Ag overlap (and thus hybridize) with the top of their valence bands, and thus giving rise to a number of interesting anomalies involving negative spin orbit splittings (-0.016 eV for CuGaS$_2$, (Ref. \onlinecite{Horinaka1978}) and nonmonotonic behavior of the energy gap versus temperature.\cite{Serrano2002}

In this article we focus on the chalcopyrite CuGaS$_2$. These sulfides, CuInS$_2$  and the corresponding selenides are being considered, together with their alloys, as efficient photovoltaic materials.
From the fundamental point of view CuGaS$_2$ has received less attention than the other related chalcopyrites, a reason why here we concentrate on the physical properties of this material.
CuGaS$_2$ was first synthesized  by Hahn \textit{et al.}.\cite{Hahn1953} They also determined by X-ray
diffraction  the crystal structure and the lattice parameters $a$ = $b$, $c$, and $u$ of this and 19 other related chalcopyrite compounds. Many of this parameters agree reasonably well with those determined experimentally  up to date and also with recent \textit{ab initio} calculations. CuGaS$_2$ was first found as a mineral (gallite) in Namibia and in the Kongo.\cite{Strunz1958,Strunz1959}

CuGaS$_2$ crystallizes with the chalcopyrite structure
which is closely related to that of zincblende with a slight distortion resulting from the tetrahedral bonding of the latter.\cite{Hahn1953}

\section{Theoretical Details}\label{SecTheory}
The calculations reported here concern the lattice parameters of  CuGaS$_2$, the Raman, ir active, and silent $k$ = 0 phonons , their Gr\"uneisen parameters and the phonon dispersion relations, the densities of phonon states (DOS) (including the projections on the vibrations of the 3 component atoms), and the optically active densities of two-phonon states. In addition, we present \textit{ab initio} calculations of the elastic constants and the bulk moduli $B_0$  and $B_0$$'$ . Because of the large number of phonon at $k$ = 0 (24) we surmised that for a calculation of the volume expansion coefficient vs. $T$ a BZ sampling using only the 24 Gr\"uneisen parameters of these phonons (at $k \approx$ 0) would yield a reasonable approximation to the scarce experimental results available. This conjecture turned out to be correct. Finally, we used the calculated phonon density of states (PDOS) to evaluate the specific heat at constant volume (and the expansion coefficient to evaluate the measured constant pressure counterpart). These calculations were performed with the natural isotopic abundance of the constituents of CuGaS$_2$  and also for crystals composed of isotopically pure atoms.

\begin{figure}[htp]
\includegraphics[width=8cm ]{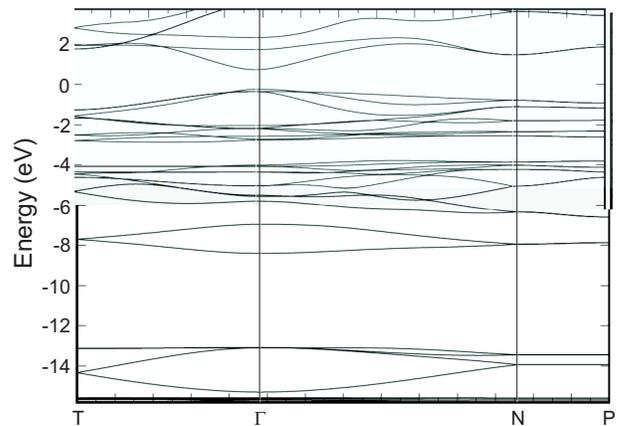}
\caption{Electronic band structure of CuGaS$_2$ calculated with the VASP code using the LDA exchange-correlation potential without SO interaction. The four lowest bands involve mainly 3$s$ electrons of S. The bands between 0 and -8 eV correspond to twenty 3$d$ electrons of Cu and twenty four 3$p$ of S. Notice that the calculated energy gap (1.1 eV) is much smaller than the experimental one (2.4 eV, Ref. \onlinecite{Jaffe1984}), reflecting the so-called "gap-problem".\cite{Godby1988} The notation of the special points in the Brillouin zone (T - $\Gamma$ - N - P) is identical to that of Ref. \onlinecite{Shileika1973}.}
\label{Fig1}
\end{figure}

The calculations were based on \textit{ab initio} electronic band structure determinations using density functional theory with either the ABINIT  or the VASP  code.\cite{Gonze2002,Gonze2009,Kresse1996,Kresse1999}
In the ABINIT calculation, normalized pseudo potentials were generated by
using the Fritz Haber Institute code  with
a valence electron configuration of 3$d^{10}$ 4$s^1$ for Cu, 3$d^{10}$ 4$s^2$ 4$p^1$ for Ga
and 3$s^2$ 3$p^4$ for S.\cite{Fuchs1999} The wave function was expanded
in plane wave up to an energy cutoff of 40 Ha and the Brillouin zone was
sampled by using the Monkhorst-Pack method with a 6 $\times$ 6 $\times$ 6 \textbf{k}-point grid.
While most electronic calculations were performed without spin-orbit (SO) interaction, in order to reveal
possible effects of this interaction on the lattice properties as well as the negative sign of the SO splitting at the top of the valence band,  a few band structure calculations with the VASP code including SO interaction were also performed. The effects of this interaction on the lattice parameters and dynamics was found to be insignificant. We display in Fig. \ref{Fig1} the electronic band structure calculated with the VASP code using the LDA exchange- correlation potential without SO interaction. The details of the \textit{ab initio} electronic band structure calculations have been given in Ref. \onlinecite{Cardona2010b}.

This band structure is similar to those reported recently by Soni \textit{et al.} and Brik  using a GGA exchange-correlation potential.\cite{Soni2010,Brik2009} Notice that in our case and that of Soni  \textit{et al.} the direct gap (1 - 2 eV) is considerably smaller than the experimental one (2.43 eV), a rather general property of LDA calculations sometimes referred to as the "gap problem".\cite{Godby1988}
Brik brought the calculated gap to agree with the experimental one by using the so-called 'scissors operator'.\cite{Brik2009}

An interesting feature of the band structures of Fig. \ref{Fig1} and Ref. \onlinecite{Soni2010} is the fact that the bands which correspond to the 3$d$ electrons of the copper overlap with the 2$p$ bands of sulfur. This peculiarity results in a negative SO splitting at the top of the valence bands (-0.016 eV) (Ref. \onlinecite{Bell1975}),  an anomalous sign also observed in CuCl (Ref. \onlinecite{Shindo1945}), ZnO (Ref. \onlinecite{Carrier2004}), and $\beta$-HgS (Ref. \onlinecite{Carrier2004,Cardona2009a}). Here we shall no longer discuss electronic properties of chalcopyrites and concentrate on the lattice parameters, phonons and thermodynamic properties.

The reader may wonder why we use two different DFT codes. The reason is that we are familiar with both codes and, while we know that they lead to similar results in the case of monoatomic and binary crystals, it is not obvious that they also will do so for more complicated structures. This was shown to be so in the case of cinnabar ($\alpha$- HgS\cite{Cardona2009a}), with three atoms per PC. Here we examine chalcopyrites, with 8 atoms per PC.

The wave function representation for the two codes is different as well as
the methodology  e. g. to calculate the
phonon vibrational spectra. Therefore, the use of both codes is
complementary.  In some cases, the ABINIT code
is more computationally demanding but the precision can be increased. Therefore,
for some properties, such as  the phonon spectra, we have employed the ABINIT implementation where density functional perturbation theory was
used.\cite{Gonze1997a,Gonze1997b}

\section{Experimental Details}\label{SecExperimental}

The CuGaS$_2$ crystals investigated here were grown by a vapor phase transport technique using iodine as transport agent.  The isotopically nearly pure (99.5\% $^{34}$S, 99.9\% $^{63}$Cu and 99.6\% $^{72}$Ga) elements were purchased from Trace Science International, Ontario, Canada.

For the single crystal X-ray structure determination
CuGaS$_2$ crystals  were selected under
the polarization microscope for crystallographic investigations. As many of
the crystals showed twinning according to $\bar{\textbf{1}}$, with ${\textbf{1}}$ being the identity operation,
a small cuboid fraction (0.24 $\times$ 0.20 $\times$ 0.14 mm$^3$) of a yellow transparent crystal was oriented on a
four-circle diffractometer using graphite-monochromatized Ag-K$\alpha$ radiation
(CAD 4, Enraf Nonius, Delft, The Netherlands). The lattice parameters of the
tetragonal lattice (space group {\sl I}\=42{\sl d}, No. 122) were refined
from 25 centered high-indexed reflections to {\sl a} = 5.3512(6) and {\sl c}
= 10.478(3) \AA. Bragg intensities in three octants of the Ewald sphere (-9
$\leq$ {\sl h, k} $\leq$ 9, -18 $\leq$ {\sl l} $\leq$ 18, 3.4$^{\rm o}$ $\leq$ $\vartheta$
$\leq$ 29.9$^{\rm o}$, {\sl R}$_{int}$ = 0.0648 after merging) were collected in the
$\omega$-$2\vartheta$ data acquisition mode and
corrected for Lorentz, polarization and absorption effects. The structure
refinement was performed with full matrix least-squares cycles on {\sl
F}$^2$ ({\sl R}1 = 0.0472 for {\sl I} $\geq$ 2 $\sigma${\sl I} and 0.0832
for all {\sl I}, {\sl wR}2 = 0.1178 for {\sl I} $\geq$ 2 $\sigma${\sl I} and
0.1350 for all).\cite{FIZ2010}

Temperature dependent  lattice parameters and the thermal expansion coefficients were determined by powder X-ray diffraction
($\lambda$ = 1.54 \AA ) on crushed crystals grown from elements with the natural isotopic abundance.

The heat capacities were measured on samples of typically $\sim$ 20 mg between 2 and 280 K with a physical property measurement system (Quantum Design, San Diego, CA) as described in detail in Ref. \onlinecite{Serrano2006}.
Between room temperature and  1100 K the heat capacities of a $\sim$100 mg polycrystalline sample were determined with a
DSC 404 F1 Pegasus differential scanning calorimeter (heating rate 20 K/min) with the sample kept in an argon atmosphere.
\cite{Netzsch} Up to 1100 K a reduction of the sample mass was not observed.

\section{Results and Discussion}\label{Results}

\subsection{Crystal Structure}

The available structural parameters of CuGaS$_2$ exhibit considerable dispersion, especially the $x$ positional parameter of the S atoms which is very close to 1/4 (in the forthcoming called $u$) has so far been determined with limited reliability and found to be larger than 1/4. Additionally, $u$ has been found to vary non-monotonically from CuAlS$_2$, via CuGaS$_2$  to CuInS$_2$.\cite{Jaffe1984} In order to increase the accuracy of $u$ we redetermined the crystal structure of CuGaS$_2$ using high-quality single crystals and up-to-date X-ray diffraction techniques which allow us to decrease the experimental error in $u$ by a factor of $\sim$7.
Table \ref{Table1} summarizes the results of our crystal structure redetermination.

\begin{table*}[htp]
\begin{center}
\caption[]{Summarized results of the crystal structure  determination ($T$ = 293(2) K) of CuGaS$_2$ (space group {\it I}\=42{\it d}, No. 122). Standard deviations of the last digit
are given in parentheses. The lattice parameters were determined to be $a$ =  5.3512(6) \AA , $c$ = 10.478(3) \AA\ resulting in a unit cell volume of 300.03(9) \AA$^3$ which contains 4 formula units ($Z$= 4). Further experimental details are deposited under the no. CSD-422615
at the Fachinformationszentrum Karlsruhe.\cite{FIZ2010}}
\begin{tabular}{lllllll}\hline\hline
\multicolumn{7}{l}{Fractional atomic coordinates and equivalent isotropic displacement parameters (\AA$^2$)} \\
\hline
Atom & Wyckoff-Nr. & {\it x} & {\it y} & {\it z} & {\it U}$_{\rm equiv}$ & \\
Ga & 4{\sl a} & 0 & 0 & 0   & 0.0078(4) & \\
Cu & 4{\sl b} & 0 & 0 & $\frac{1}{2}$  & 0.0159(4) & \\
S  & 8{\it d} & 0.2437(5) & $\frac{1}{4}$ & $\frac{1}{8}$ & 0.0085(4) & \\
\hline
\multicolumn{7}{l}{Anisotropic displacement parameters $U_{ij}$(\AA$^2$)} \\
\hline
Atom & {\sl U}$_{11}$ & {\sl U}$_{22}$ & {\sl U}$_{33}$ & {\sl U}$_{12}$ & {\sl U}$_{13}$ & {\sl U}$_{23}$ \\
Ga & 0.0080(4) & 0.0080(4) & 0.0076(5) & 0 & 0 & 0 \\
Cu & 0.0155(5) & 0.0155(5) & 0.0165(6) & 0 & 0 & 0 \\
S  & 0.0113(11)& 0.0051(10)     & 0.0092(6) & 0 & 0 & -0.0016(5) \\
\hline
\hline
\end{tabular}
\end{center}

\label{Table1}
\end{table*}

The positional parameter $u$ is clearly smaller than 1/4 ($\sim$ -2.5\%) and decreases monotonically throughout the series CuAlS$_2$ - CuGaS$_2$ - CuInS$_2$.
Table \ref{Table2}  summarizes the experimental and calculated structural parameters of CuXS$_2$ (X = Al, Ga, In) as obtained from our calculations and as available in the literature.

\begin{table}[b]
\caption{Comparison of the results of VASP-NOSO-LDA calculations performed for the three isostructural compounds CuXS$_2$ (X = Al, Ga, In) with experimental data obtained in this work and in the literature. }
\begin{tabular}{lcccccc}\hline\hline
parameter & \multicolumn{2}{c}  {CuAlS$_2$$^a$}& \multicolumn{2}{c} {CuGaS$_2$}& \multicolumn{2}{c} {CuInS$_2$$^a$} \\
\hline
  & VASP & exp & VASP & exp & VASP & exp\\
$a$ = $b$ (\AA) & 5.2055 & 5.326 & 5.226 & 5.3512(6) & 5.482 & 5.5221\\
$c$ (\AA) & 10.3765  & 10.436 & 10.380 & 10.478(3) & 10.9301 & 11.1043\\
$u$ & 0.2505 & 0.271 & 0.2450 & 0.2437(5) & 0.21651 & 0.2145\\
\hline\hline
\multicolumn{7}{l} {$^a$ Ref. \onlinecite{Zunger1983} and average  values of references therein.}
\label{Table2}
\end{tabular}
\end{table}

Since the available  calculations did not include SO splitting we performed, for comparison, calculations  with SO splitting. They are displayed in Table \ref{Table3} together with a VASP-NOSO calculation carried out by Chen \textit{et al.}\cite{Chen2007} using a GGA DFT for the exchange-correlation hamiltonian.

\begin{table}[htp]
\caption{Lattice parameters, fractional $x$ atomic coordinate of sulphur ($u$), cell volume and bulk moduli, $B_0$ and $B_0$$'$ as obtained from our \textit{ab initio} calculations either with spin-orbit coupling (SO) or without (NOSO). }
\begin{tabular}{lcccccc}\hline\hline
code & $a$ (\AA) & $c$ (\AA) & $u$ & $V_{\rm cell}$ (\AA$^3$)  & $B_0$ (GPa) & $B_0$$'$ \\
\hline
ABINIT-NOSO  & 5.262 & 10.45197& 0.244 & 289.4 & 93.3 & 4.7   \\
VASP-NOSO-LDA & 5.226 & 10.380 & 0.2450 & 283.5 & 94.3 & 4.5 \\
VASP-SO-LDA & 5.2258 & 10.3818 & 0.2450 & 283.5 & 91.9 & 5.1\\
VASP-NOSO-GGA$^a$ & 5.3700 & 10.643 & 0.2491 & 306.9 & 85 & 4.7\\
\hline\hline
\multicolumn{7}{l} {$^a$ Ref. \onlinecite{Chen2007}}
\label{Table3}
\end{tabular}
\end{table}

\subsection{Elastic Properties}

The ABINIT code used for our calculations provides also the stiffness constants, $C_{ij}$,\cite{Hamann2005} six independent ones in the case of the chalcopyrites (Table \ref{Table4}).

\begin{table}[htp]
\caption{Comparison of the stiffness constants obtained from ABINIT calculations performed for the three isostructural compounds CuGaS$_2$, CuGaSe$_2$ and AgGaS$_2$  with experimental data obtained in this work and in the literature.}
\begin{tabular}{lcccccc}\hline\hline
parameter & \multicolumn{2}{c}  {CuGaS$_2$}& \multicolumn{2}{c} {CuGaSe$_2$}& \multicolumn{2}{c} {AgGaS$_2$} \\
  & calc & exp & calc & exp & calc & exp\\
  \hline
$C_{11}$ (GPa) & 132.23  & -      & 112.2$^a$  & -        & 85.3$^b$ & 86.5$^b$\\
$C_{12}$ (GPa) & 78.4    & -      & 66.4$^a$   & -       & 52.4$^b$ & 56.0$^b$\\
$C_{13 }$ (GPa)& 79.4    & -      & 68.1$^a$   & -       & 59.9$^b$ & 59.6$^b$\\
$C_{33}$ (GPa) & 144.1   & -      & 113.2$^a$  & -       & 76.2$^b$ & 75.1$^b$\\
$C_{44}$ (GPa) & 56.1    & -      & 48.4$^a$   & -       & 32.4$^b$ & 24.9$^b$\\
$C_{66}$ (GPa) & 56.3    & -      & 48.5$^a$   & -       & 36.1$^b$ & 31.4$^b$\\
$B_0$          & 96.4    & 94$^c$, 96$^d$, 97$^e$ &  83$^a$    & 102$^a$ & 60$^d$, 62.3$^b$ & 72.2$^b$\\
$B_0$$'$   (GPa) & 4.5     & 6.3$^d$, 4$^e$     &  -         & -       & 4.2$^b$  & 4$^b$\\
\hline\hline
\multicolumn{7}{l} {$^a$ Ref. \onlinecite{Parlak2006}}\\
\multicolumn{7}{l} {$^b$ Ref. \onlinecite{Hou2010}: average values of references therein.}\\
\multicolumn{7}{l} {$^c$ Ref. \onlinecite{Bettini1975}}\\
\multicolumn{7}{l} {$^d$ Ref. \onlinecite{Werner1981}}\\
\multicolumn{7}{l} {$^e$ Ref. \onlinecite{Tinoco1994}}\\
\label{Table4}
\end{tabular}
\end{table}

We have not been able to retrieve experimental values for these from the literature. In Table \ref{Table4} we display for comparison data for CuGaSe$_2$ and AgGaS$_2$. For the latter experimental data are also quoted. We have also included bulk modulii, $B_0$, and their pressure derivative, $B_0$$'$, for all three compounds. Except for $B_0$$'$, all elastic parameters undergo a monotonic decrease through the series CuGaS$_2$, CuGaSe$_2$ to AgGaS$_2$, which is likely to be related to the corresponding increase of the lattice parameters.

\begin{figure}[htp]
\includegraphics[width=8cm ]{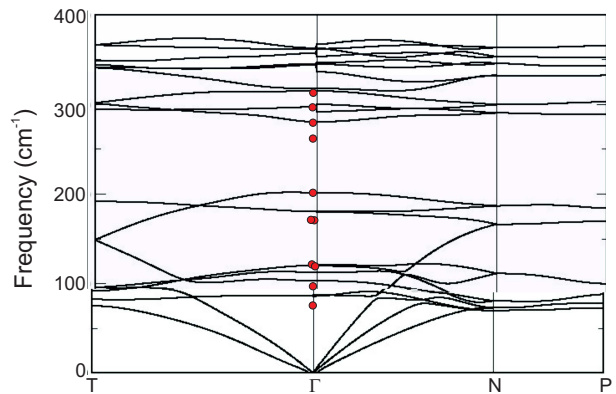}
\caption{(color online) Phonon dispersion relations of CuGaS$_2$ as calculated with the ABINIT-LDA code. The notation of the special points in the Brillouin zone (T - $\Gamma$ - N - P) is identical to that of Ref. \onlinecite{Shileika1973}. The (red) dots represent some of the available experimental ´frequencies (see Table \ref{Table5}).}
\label{Fig2}
\end{figure}

The calculated phonon dispersion relations are shown in Fig. \ref{Fig2} in the reduced Brillouin zone. The $\Gamma$ - T direction corresponds to [001] whereas $\Gamma$ - N corresponds to [110].\cite{Shileika1973} For comparison, we have added a few points obtained from Raman and infrared spectroscopy measurements given in more detail in Table \ref{Table5} where they are also compared with our \textit{ab initio} calculations and those of Akdo\v{g}an \textit{et al.}.\cite{Carlone1980,Akdogan2002}

\begin{table}[htp]
\caption{Calculated and measured phonon frequencies of CuGaS$_2$ (in cm$^{-1}$)   at the center of the BZ. Also, averaged experimental values reported in about 9 different publications (see Akdo\v{g}an \textit{et al.}\cite{Akdogan2002}, Table 4). The values of the corresponding Gr\"uneisen parameters, $\gamma$, calculated by us are also listed and compared with measurements.}
\begin{tabular}{ccccccc}\hline\hline
irred.
&	Akdo$\breve{\rm g}$an$^a$	& ours
&	ours
&	 & $\gamma$	&  $\gamma$\\
reps. & theory & ABINIT & VASP & exp.$^a$ & ABINIT & exp$^b$\\
\hline
$\Gamma_1$     (A1)	& 290.0 &	292	& 316.8 &	312	& 2.04	& 1.5\\
$\Gamma_2$     (A2)&    	341.7 &	314 &	345.0 &	silent &	1.62 &\\	
$\Gamma_2$       (A2)&	268.7 &	345 &	314.0 &	silent &	 1.68 &\\	
$\Gamma_3$      (B1)	&    329.7 &	361 &	361.2 &	372 &	1.62 &\\	
$\Gamma_3$       (B1)&	195.7 &	203 &	202.7 &	202 &	1.31 &	2.6\\
$\Gamma_3$       (B1)&	 99.0 &	114 &	114.0 &	117 &	0.47 & \\	
$\Gamma_4^{\rm LO}$  (B2) &	384.7 &	381.8 &	361.9 &	387 &	1.54&	1.5\\
$\Gamma_4^{\rm LO}$ (B2) &	234.3 &	292.0&	280.2 &	277 &	2.04 &	1.3\\
$\Gamma_4^{\rm LO}$  (B2)&	 99.0 &	104.7&	103.6	& 95 &	0.10 & \\	
$\Gamma_4^{\rm TO}$  (B2)&	354.0 &	365.4 &	361.9 & 	366	 &1.54 &	1.4\\
$\Gamma_4^{\rm TO}$  (B2)&	234.3 &	280.2&	280.2	&261&	2.07 &	1.3\\
$\Gamma_4^{\rm TO}$   (B2)&	 98.7	&103.6 &	103.6 &	 95 &	0.10 &	2.0\\
$\Gamma_5^{\rm LO}$   (E)&	367.0&	361.2 &	352.6 &	387 &	1.58 &	1.3\\
$\Gamma_5^{\rm LO}$    (E)&	327.7&	344.9 &	335.5 &	350 &	2.00 &\\	
$\Gamma_5^{\rm LO}$    (E)&	240.6 &	292.1 &	292.1 &	277 &	2.03 &	1.1\\
$\Gamma_5^{\rm LO}$    (E)&	162.3 &	180.7 &	180.7 &	169 &	1.85 &	1.5\\
$\Gamma_5^{\rm LO}$     (E)&	116.6 &	119.4 &	119.4 &	148 &	-1.04 &	0.8\\
$\Gamma_5^{\rm LO}$     (E)	& 83.3 &	 86.2&	 86.1 &	 74 &	-1.04 &	 -0.80\\
$\Gamma_5^{\rm TO}$     (E)	 &345.7&	352.6 &	352.6 &	365 &	1.65 &	1.5\\
$\Gamma_5^{\rm TO}$    (E) &	313.3 &	335.5 &	335.5	&331& 	 2.04&	1.2\\
$\Gamma_5^{\rm TO}$    (E)&	236.0 &	292.1&	292.1 &	292&	2.07 &	1.41\\
$\Gamma_5^{\rm TO}$    (E)&	161.6 &	180.7&	180.7 &	162&	1.86 &	0.8\\
$\Gamma_5^{\rm TO}$    (E)&	116.7 &	119.4&	119.4 &	115 &	-0.08 &\\	
$\Gamma_5^{\rm TO}$    (E)&	 83.3 &	 86.2	& 86.1 &	75 &	-1.04 &	 -0.80\\
\hline\hline
$^a$ Ref. \onlinecite{Akdogan2002}\\
$^b$ Ref. \onlinecite{Carlone1980}\\
\label{Table5}
\end{tabular}
\end{table}

\begin{figure}[htp]
\includegraphics[width=8cm ]{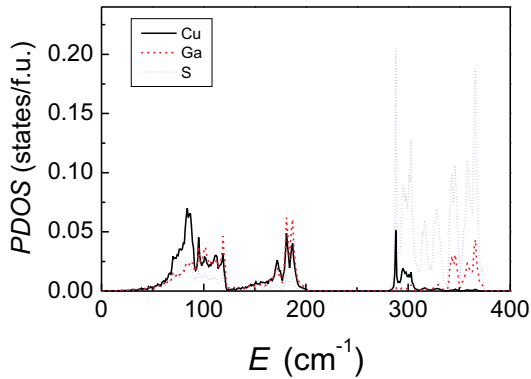}
\caption{(color online) Phonon density of states (PDOS) projected on the three constituent atoms. Note the gap between the contributions of (essentially) Cu and Ga (below 200 cm$^{-1}$) and those  above $\sim$ 280 cm$^{-1}$ which are essentially S like. There is considerable S like weight below $\sim$120 cm$^{-1}$.
The sum of these partial  densities of states has been used to calculate the temperature dependence of heat capacities.}
\label{Fig3}
\end{figure}

Figure \ref{Fig3} displays the phonon densities of states corresponding to the motion of the  three constituent atoms calculated from the dispersion relations shown in Fig. \ref{Fig2}.
As expected, the low-frequency band
0 - 100 cm$^{-1}$ corresponds mainly to Cu and Ga vibrations whereas the S-like contributions are mainly above the $\sim$280 cm$^{-1}$-gap. we note that there are also some S-like contributions below $\sim$120 cm$^{-1}$ originating from
Cu - S vibrations.
The partial densities of states are e.g. useful for calculating the effect of isotope disorder on the phonon linewidths.\cite{Serrano2004a}

We have not found in the literature second-order Raman spectra of CuGaS$_2$ which would correspond to the sum and difference spectra of Fig.\ref{Fig2}.
Nevertheless, it is possible to establish a correspondence between the calculated two-phonon Raman spectra of CuGaS$_2$ and (Fig.\ref{Fig3}) and the measured ones of $\beta$-ZnS shown in Fig. 1 of Ref. \onlinecite{Serrano2004b}.
We present here the calculated sum and difference densities of states of CuGaS$_2$ in the hope that they will help in interpreting measured spectra when they become available for the CuGaS$_2$ or other chalcopyrite compounds.

\begin{figure}[htp]
\includegraphics[width=8cm ]{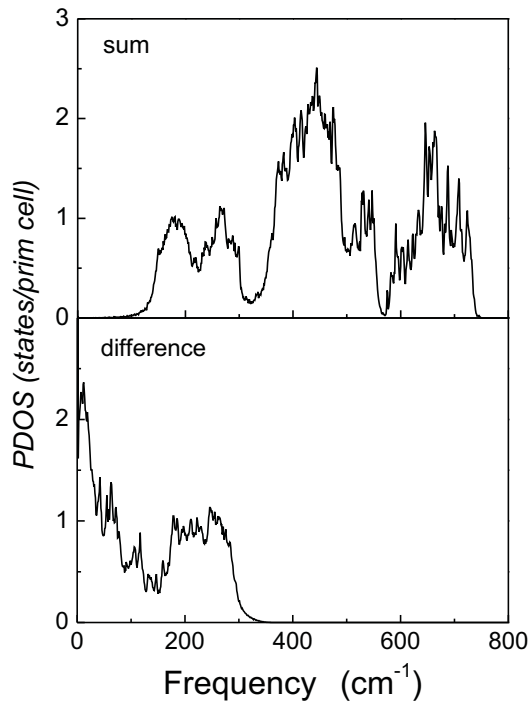}
\caption{Sum (upper panel) and difference (lower panel) phonon densities of states of CuGaS$_2$ as calculated from the dispersion relations shown in Fig. \ref{Fig4}.}
\label{Fig4}
\end{figure}

Table \ref{Table5} also lists the calculated and experimental Gr\"uneisen parameters of $\Gamma$-point phonons. As usual, most of the the Gr\"uneisen parameters are positive except some at the lowest frequencies.

The effect of lattice vibrations on the  volume $V_0 (T)$ of a (cubic) crystal can be expressed in terms of the
mode Gr\"uneisen parameters $\gamma_{\textbf{q}j}$ and the mode frequency as

\begin{equation}
\frac{\Delta V_0(T)}{V_0}=\frac{\hbar}{B_0 V}\sum_{\textbf{\textit{q}}j} \gamma_{\textbf{\textit{q}}j}\,\omega_{\textbf{\textit{q}}j} [n_B(\omega_{\textbf{\textit{q}}j}) + \frac{1}{2}],
\label{Eq1}
\end{equation}

where $n_B$ is the Bose-Einstein factor

\begin{equation}
n_B(\omega_{\textbf{\textit{q}}j})= [e^{\hbar \omega_{\textbf{\textit{q}}j}/k_{\rm B}T} -1]^{-1},
\label{Eq2}
\end{equation}

and $V$ and $B_0$ are the volume and the bulk modulus of the crystal, respectively.
\cite{Cardona2005}

Because of the large number of phonon bands we surmised that
the Gr\"uneisen parameters of the phonons at $\Gamma$ suffice, to a first approximation, for the evaluation of the volume thermal expansion coefficient, $\alpha_V$= (1/$V_0$)\,\,d$V_0(T)$/d\,$T$, to include in the summation in Eq. (\ref{Eq3}) the phonons and their Gr\"uneisen parameters at the $\Gamma$-point, where $V_0$ represents the volume of the primitive cell. As usual in semiconductors, negative values of $\gamma_{qj}$ lead to negative thermal expansion coefficients at low temperatures.\cite{Debernardi1996}
  The temperature dependence of $V_0$ is given by

\begin{equation}
\frac{\Delta V_0(T)}{V_0}=\frac{\hbar}{B_0 V}\sum_{\textbf{\textit{}}j} \gamma_{\textbf{\textit{}}j}\,\omega_{\textbf{\textit{}}j} [n_B(\omega_{\textbf{\textit{}}j}) + \frac{1}{2}],
\label{Eq3}
\end{equation}

where $\omega_j$ and $\gamma_j$ are the zone-center phonon frequencies and the Gr\"uneisen parameter, respectively.

The volume thermal expansion coefficient can also be obtained from the variation of the entropy, $S(P,T)$, with  pressure via the thermodynamic relationship

\begin{equation}
\alpha_V(T) = - \frac{1}{V}\,\left(\frac{\partial\,S(P,T)}{\partial\,P}\right)_T.
\label{Eq3B}
\end{equation}

\begin{figure}[htp]
\includegraphics[width=8cm ]{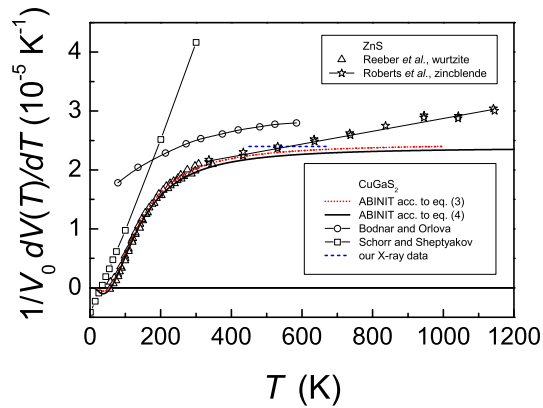}
\caption{(color online) (red) dotted line: Temperature dependence of the volume thermal expansion coefficient of CuGaS$_2$  as obtained from the zone-center Gr\"uneisen parameters $\gamma$ and frequencies as given in Table \ref{Table5} using eq. (\ref{Eq3}). (black) solid line thermal expansion coefficient from the pressure dependence of the entropy (eq. (\ref{Eq3B})). Literature  data by Reeber \textit{et al.} and Roberts \textit{et al.} and our X-ray results are also given.\cite{Reeber1967,Roberts1981}}
\label{Fig5}
\end{figure}

We calculated within the ABINIT code the entropy at ambient pressure and for  pressures of 0.4 and 1 GPa and took the numerical derivatives.
In Fig. \ref{Fig5} we compile these results and the
the results of our calculations using the $\Gamma$-point phonon frequencies and the mode Gr\"uneisen parameters summarized in Table \ref{Table5} (and its temperature derivative), i.e. the volume thermal expansion coefficient, $\alpha_V(T)$ = ($1/V_0$)\,\,$d\,V(T)/d\,T$,   with literature data by Bodnar \textit{et al.}, Schorr \textit{et al.}, and with our high temperature X-ray data. For comparison we also display the volume thermal expansion coefficient of Reeber and Powell \textit{et al.}.\cite{Bodnar1983,Schorr2008,Reeber1967}

The volume thermal expansion coefficients obtained from the $\Gamma$-point phonon frequencies and from the pressure-derivative of the entropy are in good agreement, thus justifying the approximation used in eq. (\ref{Eq3}).
The results of our calculations clearly reveal a negative thermal volume expansion coefficient at low temperatures as has been found experimentally by Schorr and Scheptyakov.\cite{Schorr2008} However, at high temperatures our calculated data deviate markedly from the experimental findings, they are about 15\% lower than the data by Bodnar and Orlova obtained by X-ray diffraction experiments.\cite{Bodnar1983} However, there still appears to be some scatter in the experimental data. While Bodnar \textit{et al.} report a room temperature volume thermal expansion coefficient of 24$\times$10$^{-6}$K$^{-1}$ close to 26.5$\times$10$^{-6}$K$^{-1}$ obtained by Yamamoto \textit{et al.}, Malsagov \textit{et al.} found a room temperature value of only 19.2$\times$10$^{-6}$K$^{-1}$, close to the results of our calculations.\cite{Bodnar1983,Yamamoto1979,Malsagov1986}
Our calculations also agree rather well with our high-temperature X-ray diffraction data carried out on a polycrystalline sample prepared from the same crystals used for the heat capacity experiments (see below).
The temperature dependence of the volume thermal expansion coefficient for CuGaS$_2$ is expected to be similar to that of the isobaric and isoelectronic
ZnS. This is indeed the case; up to room temperature our calculations coincide remarkable well with volume thermal expansion data for ZnS obtained by Reeber and Powell.\cite{Reeber1967} Above room temperature Roberts \textit{et al.} observed a linear increase of the volume thermal expansion coefficient of zincblende ZnS at a rate of 1.16$\times$10$^{-8}$K$^{-2}$ which is not found in our calculations for CuGaS$_2$.\cite{Roberts1981}
This linear increase of the volume thermal expansion coefficient of wurtzite ZnS and zincblende ZnS has  recently also been obtained by  \textit{ab initio} calculations.\cite{Wang2006}

\subsection{Heat Capacity}

The phonon density of states displayed in Fig. \ref{Fig3} allows calculation of the free energy
$F(T)$ by using the expression:

\begin{equation}
F(T)=-\int_0^\infty (\frac{\hbar\omega}{2} +  k_{B}T
ln[2n_B(\omega)]) \rho(\omega)d\omega\label{Eq4}
\end{equation}

and of the specific heat at constant volume by taking the second derivative of the free energy

\begin{equation}
C_V =-T\left(\frac{\partial^2 F}{\partial T^2}\right)_V.\label{Eq5}
\end{equation}

In Eq. (\ref{Eq4}), $k_{B}$ is the Boltzmann constant, $n_B$ the
Bose-Einstein factor, and $\rho(\omega)$ the phonon density of states. The high frequency cutoff of the latter defines the
upper limit of integration in Eq. (\ref{Eq4}).

The difference between the calculated heat capacity at constant volume, $C_V$, and that  at constant pressure, $C_P$, (the quantity that is experimentally obtained) is related to the volume thermal expansion coefficient, $\alpha_V$,  according to

\begin{equation}
C_P(T) - C_V(T) = \alpha_v^2(T) \cdot B \cdot V_{{mol}} \cdot T,
 \label{Eq6}
\end{equation}

where    $B_0$ is the (isothermal) bulk modulus and $V_{mol}$ the molar
volume.
With $\alpha_V (T > 300 K) \approx 2.3 \times 10^{-5} K^{-1}$ the contribution to the heat capacity from thermal expansion amounts to 0.7 J/molK and 2.3 J/molK at 300 K and 1000 K, respectively.

\begin{figure}[htp]
\includegraphics[width=10cm ]{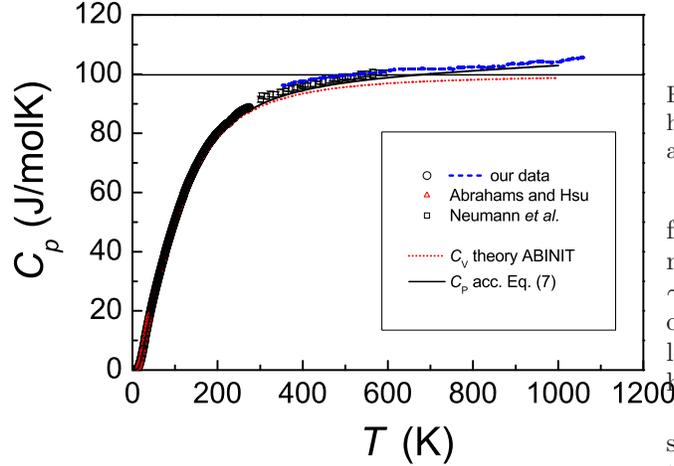}
\caption{(color online) Temperature dependence of the molar heat capacity of CuGaS$_2$. (o) and (blue) dashed line: our experimental data, (red) dotted line: results of ABINIT calculations of $C_V$, (black) solid line: $C_P$ according to Eq. (\ref{Eq6}) using our calculated volume thermal expansion coefficient $\alpha(T)$ displayed in Fig. \ref{Fig5}. Above room temperature we have added a linear increase of $\alpha(T)$ which amounted to 1.16$\times$10$^{-8}$K$^{-2}$, identical to that found by Roberts \textit{et al.} for zincblende ZnS.\cite{Roberts1981}
Literature data by Abrahams and Hsu and Neumann \textit{et al.} are also displayed.\cite{Abrahams1975,Neumann1987}
The vertical line indicates the Petit-Dulong value of 12$\times R$, where $r$ is the molar gas constant.}
\label{Fig6}
\end{figure}

\begin{figure}[htp]
\includegraphics[width=8cm ]{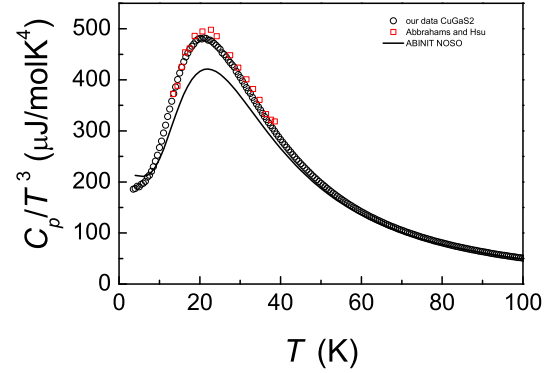}
\caption{(color online) Temperature dependence of the molar heat capacity of CuGaS$_2$. (o): our experimental data. Literature data by Abrahams and Hsu are also given.\cite{Abrahams1975}}
\label{Fig7}
\end{figure}

In the literature  heat capacity data for CuGaS$_2$ are available in the low temperature regime (13 K $\leq T \leq$ 38 K) and above room temperature up to $\sim$600 K.\cite{Abrahams1975,Neumann1987} Our data connect these temperature regimes and extend the temperature range up to 1100 K. At low temperature our results are in good agreement with the data of Abrahams and Hsu with improved resolution of the maximum in $C_P/T^3$ (cf. Fig. \ref{Fig7}), at high temperatures they connect well to the data by Neumann \textit{et al.}.
In Fig. \ref{Fig6} we display the heat capacity of CuGaS$_2$ over a temperature range from 2 K up to 1100 K together with the theoretical results based on the PDOS shown in Fig. \ref{Fig3}. To account for the contributions of the thermal expansion, becoming especially noticeable above room temperature,  we have used our data of the volume thermal expansion coefficient extended by a linear increase above room temperature of 1.16$\times$10$^{-8}$K$^{-2}$, identical to that found by Roberts \textit{et al.} for zincblende ZnS.\cite{Roberts1981} as displayed in Fig. \ref{Fig5} and used Eq. (\ref{Eq6}) to calculate $C_P$.

In some of our previous works\cite{Romero2008,Cardona2010b}  we have investigated the dependence of  $C_p/T^3$ on the isotopic masses of the  constituents of the compounds (up to binary compounds, so far) and  compared the experimental data with the theoretical results.
For elementary and binary compounds we have also investigated
the relationship of the logarithmic derivatives versus temperature and versus the masses of the  constituents.\cite{Romero2008}

\begin{figure}[htp]
\includegraphics[width=8cm ]{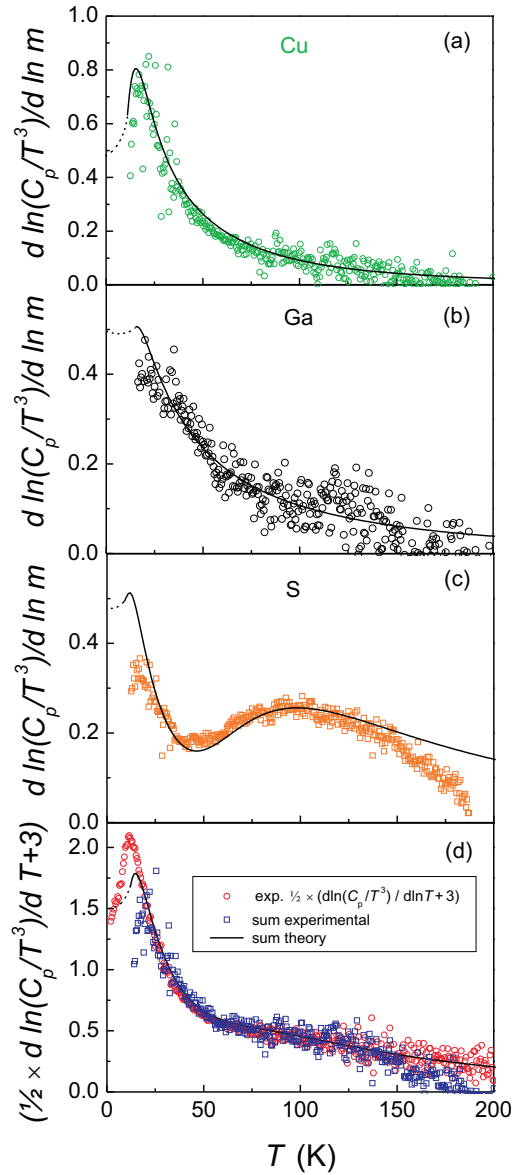}
\caption{(color online)(a - c)  Logarithmic derivatives of the specific heats, $C_p/T^3$, with respect to the atomic masses of the constituents, Cu, Ga and S (from top to bottom), respectively. The (black) solid lines represent the logarithmic derivatives calculated from our theoretical heat capacities (ABINIT).
(d) (red) o Logarithmic derivatives of the specific heat of CuGaS$_2$ (natural isotope composition) with respect to the temperature. Plotted is the quantity $\frac{1}{2}\times(d\,ln\,C_p/T^3 / d\,ln\,T +3)$ which is compared with the sum of the logarithmic derivatives with respect to the isotopic masses (blue) {\footnotesize{$\square$}}: experiment; black solid line: sum of  theory curves shown in (a - c). At very low temperatures the theoretical logarithmic derivatives have been extrapolated (dashed lines) to the values imposed for $T \rightarrow $ 0 by eq. (\ref{Eq9}-\ref{Eq11}).}
\label{Fig8}
\end{figure}

In all logarithmic derivatives we observe  a peak centered at about 20 K. For S there is an additional broad band with its maximum at about $\sim$100 K.
The maxima in the logarithmic derivatives reflect the structure of the PDOS projected on the corresponding atoms.
Clearly, the broad high energy feature visible exclusively for S originates from the PDOS with mainly S character between 280 and 380 cm$^{-1}$. The ratio of the maximum frequency (converted into Kelvin)  in the PDOS and the maximum temperature of the logarithmic derivative is  $\sim$6, a ratio which has been found in a number of previous investigations.\cite{Cardona2010b} Cu shows the highest feature in the logarithmic derivative at low temperature reflecting the high weight of the Cu projected PDOS at low energies.

We have demonstrated that there is a close relationship of the logarithmic derivatives of the heat capacities with respect to temperature and isotope mass.\cite{Cardona2007}
The straightforward extension of the relationship  the logarithmic derivatives versus temperature and versus the masses of the isotopes to the ternary compound, CuGaS$_2$ in our case,  is given by

\begin{equation}
  \frac{1}{2}\,\,(3 + \frac{d
\ln (C_{p}/T^3)}{d \ln T})=\frac{d \ln (C_{p}/T^3)}{d \ln M_{\rm Cu}} + \frac{d \ln (C_{p}/T^3)}{d \ln M_{\rm Ga}} + \frac{d \ln (C_{p}/T^3)}{d \ln M_{\rm S}}, \label{Eq8}
\end{equation}

where $M_{\rm Cu}$, $M_{\rm Ga}$, and  $M_{\rm S}$ are the masses of the three constituents, i.e. Cu, Ga,  and S, respectively, for CuGaS$_2$.

Figure \ref{Fig8} confirms the relationship of the temperature dependence of the logarithmic derivatives with respect to temperature and to the isotope masses established earlier by us is also valid for multinary (in the present case ternary) compounds.

We have demonstrated that for low temperatures, $T \rightarrow$ 0, the logarithmic derivatives are related to the ratios of the atomic mass to the molar mass according to \cite{Romero2008}

\begin{equation}
\frac{d \ln C_v/T^3}{d \ln M_{\rm Cu}} =  \frac{3}{2}\,\, \frac{M_{\rm
Cu}}{M_{\rm Cu} + M_{\rm Ga} + 2M_{\rm S}} = 0.48\\
\label{Eq9}
\end{equation}
\begin{equation}
\frac{d \ln C_v/T^3}{d \ln M_{\rm Ga}} =  \frac{3}{2}\,\, \frac{M_{\rm
Ga}}{M_{\rm Cu} + M_{\rm Ga} + 2M_{\rm S}} = 0.53\\
\label{Eq10}
\end{equation}
\begin{equation}
\frac{d \ln C_v/T^3}{d \ln M_{\rm S}} =  \frac{3}{2}\,\, \frac{2M_{\rm
S}}{M_{\rm Cu} + M_{\rm Ga} + 2M_{\rm S}} = 0.49.
\label{Eq11}
\end{equation}

In the case of CuGaS$_2$ these three ratios are, fortuitously, approximately equal to 0.5.

\section{Conclusions}\label{SectionConclus}

\textit{Ab initio} electronic band-structure techniques, especially those which use up-to-date computer codes like VASP or ABINIT, are powerful methods to investigate electronic, optical, vibronic and thermodynamic properties of crystals and the results with experimental data. Here we apply these techniques to CuGaS$_2$ which has chalcopyrite structure (space group $I$$\bar{\rm 4}$2$d$ (No. 122), two molecules per primitive cell.) more complicated than those usually dealt with. We used the ABINIT code to calculate the frequencies of Raman and IR phonons and their dispersion relations. The densities of states of one and two phonons have also been calculated. We devote the last section to present experimental data on the specific heat versus temperature of samples grown with the natural isotopic abundances and those grown with isotopically modified ones. These results are compared with \textit{ab initio} calculations. Generally, good agreement between experiment and \textit{ab initio} results is obtained. A redetermination of the crystal structure parameters is presented which decreases the discrepancies with experimental data apparent in the literature.
The peaks of the Cu and Ga mass derivatives in Fig. \ref{Fig8} are at similar temperatures. In the process of measuring AgGaS$_2$ wherein the masses of the two cations are considerably different we observe a clear difference of the low temperature phonon spectrum related to the differences of the atomic masses of Ag and Ga.

\begin{acknowledgments}
AHR has been supported by CONACYT Mexico under projects J-83247-F, Binational
Collaboration FNRS-Belgium-CONACYT and PPPROALMEX-DAAD-CONACYT.
A. M. acknowledges the financial support from the Spanish MCYT under grants MAT2010-21270-C04-03,  CSD2007-00045 and the supercomputer  resources provides by the Red Espa\~{n}ola de Supercomputaci$\acute{\rm{o}}$n.
We are grateful to Y. Pouillon and O. Castillo for valuable technical and
computational
support.
We are also indebted to E. Kotomin  for a critical reading of the manuscript.

\end{acknowledgments}

\end{document}